\documentclass[useAMS,usenatbib]{mn2e}
\usepackage{graphicx}
\title[The central WNh stars of R136]{VLT/SINFONI time-resolved spectroscopy of the
central, luminous, H-rich WN stars of R136\thanks{Based on observations collected at the European Organisation for Astronomical Research in the Southern Hemisphere, Chile, under program ID 076.D-0563, and on observations made with the Hubble Space Telescope obtained from the ESO/ST-ECF Science Archive Facility}} \author[O. Schnurr et al.]{O. Schnurr$^{1,2}$\thanks{E-mail: o.schnurr@sheffield.ac.uk}, A.-N. Chen\'e$^{2,3}$, J. Casoli$^{2,4}$, A. F. J. Moffat$^{2}$, and N. St-Louis$^{2}$
\vspace{3mm}\\ $^{1}$Dept. of Physics and Astronomy, University of Sheffield, Hicks
Building Hounsfield Road, Sheffield S3 7RH, United Kingdom\\
$^{2}$D\'ept. de Physique , Universit\'e de Montr\'eal, C. P. 6128, succ. centre-ville, Montr\'eal (Qc) H3C 3J7, and Centre de Recherche\\
en Astrophysique du Qu\'ebec, Canada\\
$^{3}$Herzberg Institute of Astrophysics, 5071 West Saanich Road, Victoria (BC) V9E 2E7, Canada\\
$^{4}$Ecole Normale Sup\'erieure, 45, rue d'Ulm, 75230 Paris C\'edex 05, France}
\begin{document}

\date{Version 18 May 2009}

\pagerange{\pageref{firstpage}--\pageref{lastpage}} \pubyear{2009}

\maketitle

\label{firstpage}

\begin{abstract}
Using the Very Large Telescope's Spectrograph for INtegral Field
Observation in the Near-Infrared (VLT/SINFONI), we have obtained
repeated AO-assisted, NIR spectroscopy of the six central luminous,
Wolf-Rayet (WR) stars in the core of the very young ($\sim1$ Myr),
massive and dense cluster R136, in the Large Magellanic Cloud
(LMC). We also de-archived available images that were obtained with
the Hubble Space Telescope's Space Telescope Imaging Spectrograph
(HST/STIS), and extracted high-quality, differential photometry of our
target stars to check for any variability related to binary motion.

Previous studies, relying on spatially unresolved, integrated, optical
spectroscopy, had reported that one of these stars was likely to be a
4.377-day binary. Our study set out to identify the culprit and any
other short-period system among our targets. However, none displays
significant photometric variability, and only one star, BAT99-112
(R136c), located on the outer fringe of R136, displays a marginal
variability in its radial velocities; we tentatively report an 8.2-day
period. The binary status of BAT99-112 is supported by the fact that
it is one of the brightest X-ray sources among all known WR stars in
the LMC, consistent with it being a colliding-wind system. Follow-up
observations have been proposed to confirm the orbital period of this
potentially very massive system.

\end{abstract}


\section{Introduction}

30 Doradus (NGC 2070) is an active star-forming, giant H\textsc{ii}
region in the Large Magellanic Cloud (LMC). At the center of 30 Dor is
located the very massive star cluster R136 (HD 38268), which is
regarded as the closest visible example of a ``super star cluster'',
excluding its Galactic clone NGC 3603, which is not surrounded by a
very massive cluster halo.

The inner arcsecond of R136, denoted R136a, is unresolvable by
conventional ground-based telescopes, and it was suspected that R136a
is in fact a single stellar object of more than 3000 M$_{\odot}$
(\citealt{Cassinelli81}, \citealt{Feitzinger80}). However, careful
ground-based work (both direct imaging: Moffat \& Seggewiss 1983;
Moffat, Seggewiss \& Shara 1985; and speckle interferometry:
\citealt{Weigelt85}) showed that R136a was not a single star.
High-resolution imaging by the \emph{Hubble Space Telescope (HST)}
confirmed that R136a indeed consists of individual, hot and luminous
stars (\citealt{Campbell92}), while the region immediately surrounding
R136 contains many dozens of massive O stars, many of them of spectral
type O3, the hottest of all known O-type stars at the time
(\citealt{MassHunt98}).

It has been theorized that high stellar densities in the cores of very
massive proto-clusters are a prerequisite for the formation of very
massive stars (\citealt{Bate02}). Moreover, there is empirical
evidence that the mass of the most massive cluster member correlates
with the total mass of the cluster (\citealt{WeidKroup06}). Therefore,
it can be expected that the most massive stars known are found in the
core regions of the most massive, and densest, unevolved clusters
known. Given its core density of $\sim10^{5}$ M$_{\odot}$pc$^{-3}$
(e.g. \citealt{Moff85}), R136a is hence a prime candidate for
harboring extremely massive stars.

It is a very remarkable fact that even the hottest and most massive
O-type stars seem not to exceed $\sim60$ M$_{\odot}$
(\citealt{Lamontagne96}; \citealt{Massey02}). More massive stars are
invariably members of a very luminous and hydrogen-rich subtype of
Wolf-Rayet (WR) stars of the nitrogen-rich sequence, the so-called
WN5-7h stars. Studies using model-atmospheres indicate that these
WN5-7h stars are not classical, core-helium burning objects, usually
identified with the WR phase, but rather core-hydrogen burning objects
on the the main sequence; their WR-like appearance is due to their
very high luminosities, $\log(\rm L/L_{\odot}) \ga 6.0$;
\citealt{deKoter97}; \citealt{CroDess98}) which drive fast stellar
winds, whose high densities give rise to the emission-line
spectrum. As has been confirmed by weighing WN5-7h stars with
Keplerian orbits in eclipsing binaries, the very high luminosities of
WN5-7h stars indeed correspond to very high masses: The present record
holder, NGC3603-A1, tips the balance at 116 M$_{\odot}$
(\citealt{S08a}).

Following the spectral classification of \citet{CroDess98}, R136a
contains three hydrogen-rich WN5h stars, and one O3f/WN6 star which is
a transition type between the hottest Of stars and the least extreme
WN5-7h stars (\citealt{Walborn86}) and thus weaker-lined than the WN5h
stars. Slightly off-centered are a cooler, evolved WN9h star (R136b),
and another WN5h star (R136c). From ground-based, spatially unresolved
(i.e. integrated), optical spectroscopy of R136a, \citet{MoffSegg83}
reported a 4.377-day WNh binary in R136a, with a diluted
radial-velocity amplitude $K \sim 38$ kms$^{-1}$; the subsequent study
by \citet{Moff85} confirmed this finding. The situation in R136a seems
thus very similar to that in NGC 3603, where from unresolved
spectroscopy of the central arcsecond, too, \citet{MoffNiem84} had
identified a 3.77-day binary among the three WN6h stars (this finding
was confirmed by Moffat et al. 1985 as well). Follow-up observations
of NGC 3603 confirmed these results (\citealt{Moff04};
\citealt{S08a}), revealing a second close binary, C, in NGC 3603, with
an orbital period $P = 8.9$ days. Hence a more detailed investigation
of the six WNh stars in R136a was warranted, since it offered the
potential to increase considerably the number of known, very massive
binary systems.

We have therefore obtained, for the first time, repeated, spatially
resolved spectroscopy of the six WNh stars in R136, in order to single
out the 4.377-day binary and to identify any further short-period
system among our targets. To search for eclipsing systems
(photospheric or atmospheric) among the stars, additional optical
\emph{HST} photometry of our target stars was extracted from publicly
available, archival imaging data.

Our study also complements that of \citet{S08b}, who had surveyed the
41 of the 47 known, late-type WN stars in the LMC according to BAT99
located outside R136a, and thereby concludes the efforts of the
Montr\'eal hot-star group to monitor every WR star in the Magellanic
Clouds to establish their binary status.

In the present paper, we describe the data acquisition and reduction
(Section 2), and the data analysis (Section 3). We discuss our
findings in Section 4, and close with the summary and conclusion in
Section 5.

\section{Observations and data reduction}

\subsection{Near-infrared spectroscopy}

Targets and their relevant properties are listed in Table
\ref{targets}, giving both the numbers from the BAT99 catalogue
(\citealt{BAT99}) and the older Radcliffe numbers
(e.g. \citealt{Feast60}).

\begin{figure}
\includegraphics[width=60mm,angle=-90,trim= 0 0 0 0,clip]{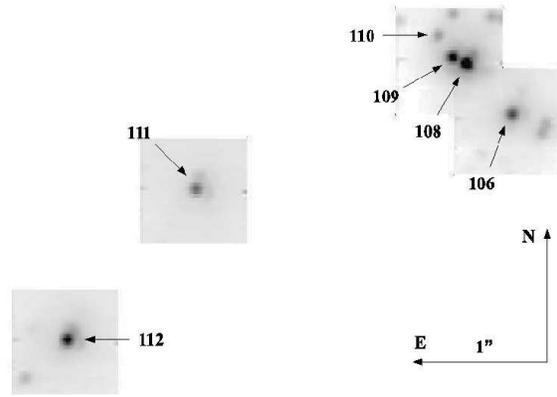}
\caption{Montage of the four $0.8'' \times 0.8''$ fields as seen on
the sky, reconstructed from the SINFONI data cube. Target stars are
identified by the BAT99 numbers. North is up and East is left.}
\label{imagecube}
\end{figure}

Observations were carried out in service mode at the Very Large
Telescope (VLT) with Unit Telescope 4 under Program-ID P076.D-0563,
between November 13 and December 5, 2005. The observations thus cover
a time span of $\sim$22 days. We obtained repeated K-band (1.95 to
2.45 $\mu$m) spectroscopy using the Spectrograph for INtegral Field
Observation in the Near-Infrared (SINFONI) (\citealt{Eisenhauer03};
\citealt{Bonnet04}) with adaptive-optics (AO) correction to obtain the
highest possible spatial and spectral resolution. BAT99-111 (R136b)
served as AO reference star. The field of view was $0.8'' \times 0.8''$
with a ``spaxel'' scale of 12.5 mas $\times$ 25 mas. Our six target
WNh stars were observed with four pointings (with R136a1, a2 and a5
all in one field), defined as telescope offsets from the AO guide star
BAT99-111.  Figure \ref{imagecube} shows a montage of the four fields
as seen on the sky, reconstructed from the SINFONI data cube. Stars
here are separated at least as well as on HST images.

\begin{table*}
\caption{Target list of the observations. For easier identification,
both BAT99 and Radcliffe numbers are given together with the spectral
types (based on optical spectra). $(b-v)$ colors and $E(b-v)$ were
directly adopted from \citet{CroDess98}, while their Johnson-$V$
magnitudes were used to estimate narrow-band $v$-band magnitudes,
applying $(v-V) \sim 0.1$ mag to take into account the contribution of
emission lines.}
\label{targets}
\begin{tabular}{ccccccc}
\hline
BAT99   & other  &  spec        & $v$   & $(b-v)$ & $E(b-v)$ & $M_{v}$\\
        & name   &  type        & mag   &  mag    &   mag    &   mag \\
\hline
106     & R136a3 &  WN5h        & 13.1 & +0.20   & 0.28     & -6.6\\
108     & R136a1 &  WN5h        & 12.9 & +0.18   & 0.28     & -6.8\\
109     & R136a2 &  WN5h        & 13.1 & +0.21   & 0.28     & -6.7\\
110     & R136a5 &  O3If*/WN6-A & 14.0 & +0.21   & 0.28     & -5.7\\
111     & R136b  &  WN9h        & 13.4 &   ...   &  ...     & -6.5\\
112     & R136c  &  WN5h        & 13.6 &   ...   &  ...     & -6.3\\
\hline
\end{tabular}
\end{table*}

Total exposure times were 150s per star and per visit, each organized
in 2 detector integration times (DITs). Given the brightness of our
targets ($K \sim$ 11--12mag) and the AO deployment, no dedicated sky
frames were taken. Other calibrations (dark and flat-field frames, and
the telluric standard star) were provided by the ESO baseline
calibration.

For most of the data reduction steps, ESO's pipeline was used
(cf. \citealt{Abuter06}). Standard reduction steps were taken. The
two-dimensional spectra produced by each illuminated slitlet were
individually extracted using \textsc{iraf}, and combined into one
wavelength-calibrated spectrum per star and visit. A main-sequence
B-type star was used for telluric corrections. To remove the B star's
Br$\gamma$ absorption which coincides with the Br$\gamma$/He\textsc{i}
$\lambda$2.166 $\mu$m emission blend of the target WNh stars, a
Lorentzian was fitted to the absorption line and subtracted from the B
star's spectrum. Residuals were very small, and proved to be harmless
in the subsequent analysis. Finally, science spectra were rectified by
fitting a low-order spline function to the stellar continuum. The
final uniform stepwidth of the spectra was 2.45 \AA/pixel, resulting
in a conservative three-pixel resolving power of $\sim$3000, and a
velocity dispersion of $\sim$33 kms$^{-1}$/pixel.


\subsection{Optical photometry}

We have de-archived optical imaging of R136 that was obtained in Cycle
8 as part of the \emph{HST} program GO-8217 (PI: Philip Massey). Using
the Space Telescope Imaging Spectrograph (STIS), 30 pairs of short
(1.1-1.3s) images were obtained at various roll angles through the
long-pass filter (F28X50LP). The image size was $28'' \times 50''$ and
the spatial scale was 0.05$''$ per pixel. For more details on the
data, we refer the reader to \citet{Massey02}. Standard pipeline data
reduction was carried out on the fly by the \emph{HST} archive
software, and no further data handling was done on our part.

\section{Data analysis and results}

\subsection{Spectroscopy}

\begin{figure*}
\includegraphics[width=100mm,angle=-90,trim= 0 57 10 38,clip]{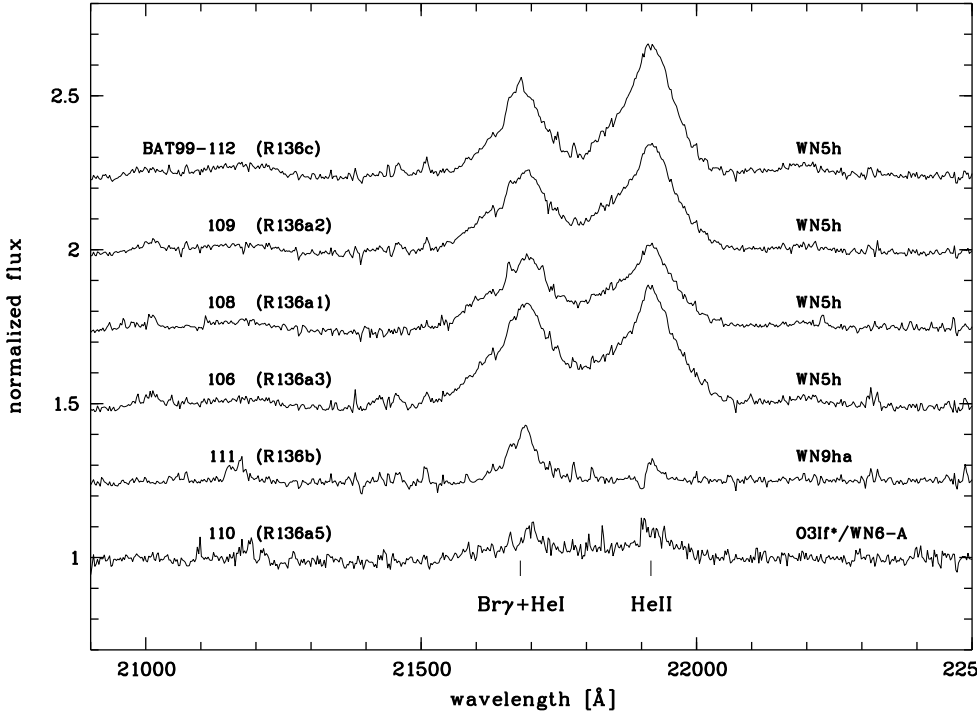}
\caption{Montage of average spectra of our six target stars, arranged
by spectral type and limited to the useful spectral region. The two
strongest emission lines, Br$\gamma$/He\textsc{i} $\lambda$2.166
$\mu$m and He\textsc{ii} $\lambda$2.188 $\mu$m are indicated. For
clarity, the spectra have been shifted by 0.25 flux units, starting with the
second from the bottom.}
\label{montage}
\end{figure*}

The mean spectra of the six WNh stars are shown in Figure
\ref{montage}. The systemic velocity $RV_{\rm sys}$ of each star was
measured by fitting Gaussians to the reasonably symmetric, top third
of the He\textsc{ii} $\lambda 2.188 \mu$m emission line (below that,
the line quickly becomes too asymmetric for this). For the WN5h stars,
which have the strongest emission lines, this was done for each
individual spectrum of the time series. The standard deviation of the
series was adopted as error, $\sigma_{\rm RV}$, and the error of the
mean of the systemic velocity was calculated as $\rm eom = \sigma_{\rm
RV}/\sqrt{N}$, with $N=9$ the number of spectra.

For the two weaker-lined stars, BAT99-110 (O3f/WN6) and 111 (WN9ha),
the fit could only be carried out for the averaged spectrum since the
S/N ratio was better. Also these two stars display obvious P Cygni
profiles in He\textsc{ii} $\lambda 2.188 \mu$m, hence two Gaussians
(one absorption, one emission) plus a linear continuum were fitted to
the entire line. The resulting errors on the fits are comparably
large, in particular for the faint O3f/WN6 star, because its very weak
lines remained buried in the noise even in the mean spectrum. Since
only one spectrum was used, no error of the mean was
calculated. Values are listed in Table \ref{scatters}.

\begin{table*}
\caption{Systemic velocities, standard deviations $\sigma_{\rm RV}$,
and errors of the mean ($\rm eom = \sigma_{\rm RV}/\sqrt{N}$) of our
program stars, obtained by fitting a single Gaussian to the
He\textsc{ii} $\lambda 2.188 \mu$m line. For BAT99-110 and 111, two
Gaussians (emission and absorption) were fitted to the P Cygni profile
to the mean spectrum; therefore, both RVs are given, but no error of
the mean was calculated. The listed RV scatter $\sigma_{\rm RV,xcorr}$
was obtained from cross-correlation for each individual star. Those
stars used as RV reference are flagged accordingly.}
\label{scatters}
\begin{tabular}{cll r@{}@{ }p{3mm}@{ }@{}r c l c l}
\hline
BAT99   & other & Spectral &  \multicolumn{3}{c}{$RV_{\rm sys}$} & eom & & $\sigma_{\rm RV,xcorr}$ & comment\\
        & name  & type & \multicolumn{3}{c}{(kms$^{-1}$)} & (kms$^{-1}$) & & (kms$^{-1}$) &\\
\hline
106     & R136a3 & WN5h  & 380   & $\pm$ & 11 & 3.7 &      & 14   &  RV reference \\
108     & R136a1 & WN5h  & 374   & $\pm$ & 19 & 6.3 &      & 19   &  RV reference \\
109     & R136a2 & WN5h  & 392   & $\pm$ & 17 & 5.7 &      & 13   &  RV reference \\
110     & R136a5 & O3f/WN6 & 288 & $\pm$ & 63 &     & em.  & 33   &  very weak P Cyg\\
        &        &       &  52   & $\pm$ & 16 &     & abs. &      &  \\ 
111     & R136b  & WN9ha & 348   & $\pm$ & 36 &     & em.  & 26   &  weak P Cyg\\
        &        &      &  157   & $\pm$ &  6 &     & abs. &      &  \\
112     & R136c  & WN5h &  389   & $\pm$ & 21 & 7.0 &      & 28   &  variable?\\
\hline
\end{tabular}
\end{table*}

All four WN5h stars display systemic velocities that are redder
(larger) than the systemic velocity expected from the LMC, $280 \pm
20$ kms$^{-1}$ (e.g. \citealt{Kim98}). This redshift is most likely of
the same nature as that reported for the optical He\textsc{ii}
$\lambda 4686$ line in other WN and O3f/WN6 stars in the LMC
(e.g. \citealt{S08b}), i.e. due to radiative-transfer effects
(\citealt{Hillier89}). The mean systemic velocity of the four WN5h
stars is $(384 \pm 4)$ kms$^{-1}$ (standard deviation), whose
dispersion is comparable to their individual errors of the mean; these
stars thus have systemic velocities which are statistically consistent
with each other.

For BAT99-110, the emission part of the Cyg profile for He\textsc{ii}
$\lambda 2.188 \mu$m surprisingly displays a significantly smaller
redshift than the He\textsc{ii} $\lambda 2.188 \mu$m emission of WN5h
stars. In fact, despite the large error, BAT99-110's systemic velocity
is in excellent agreement with that of the LMC. Either near-infrared
He\textsc{ii} emission lines of O3f/WN6 stars do not display the
intrinsic redshift of their optical counterparts (possibly because of
optical-depth effects, since their wind is thinner than that of, say,
WN5h stars) or BAT99-110 is indeed travelling at relatively high
differential velocity along the line of sight towards the observer. In
this case, it could be that 110 is not actually in the core, but
rather ejected from it towards the observer, i.e. along the line of
sight (also see below). Thus, from a point of view of cluster dynamics
and cluster evolution, it would be very interesting to carry out a
more detailed RV study of the massive stars in the periphery of R136a
to check for ejected cases (see also \citealt{Brandl07} for a more
detailed discussion of this topic).

\begin{figure*}
\begin{minipage}{165mm}
\includegraphics[width=85mm,angle=0,trim= 5 0 25
50,clip]{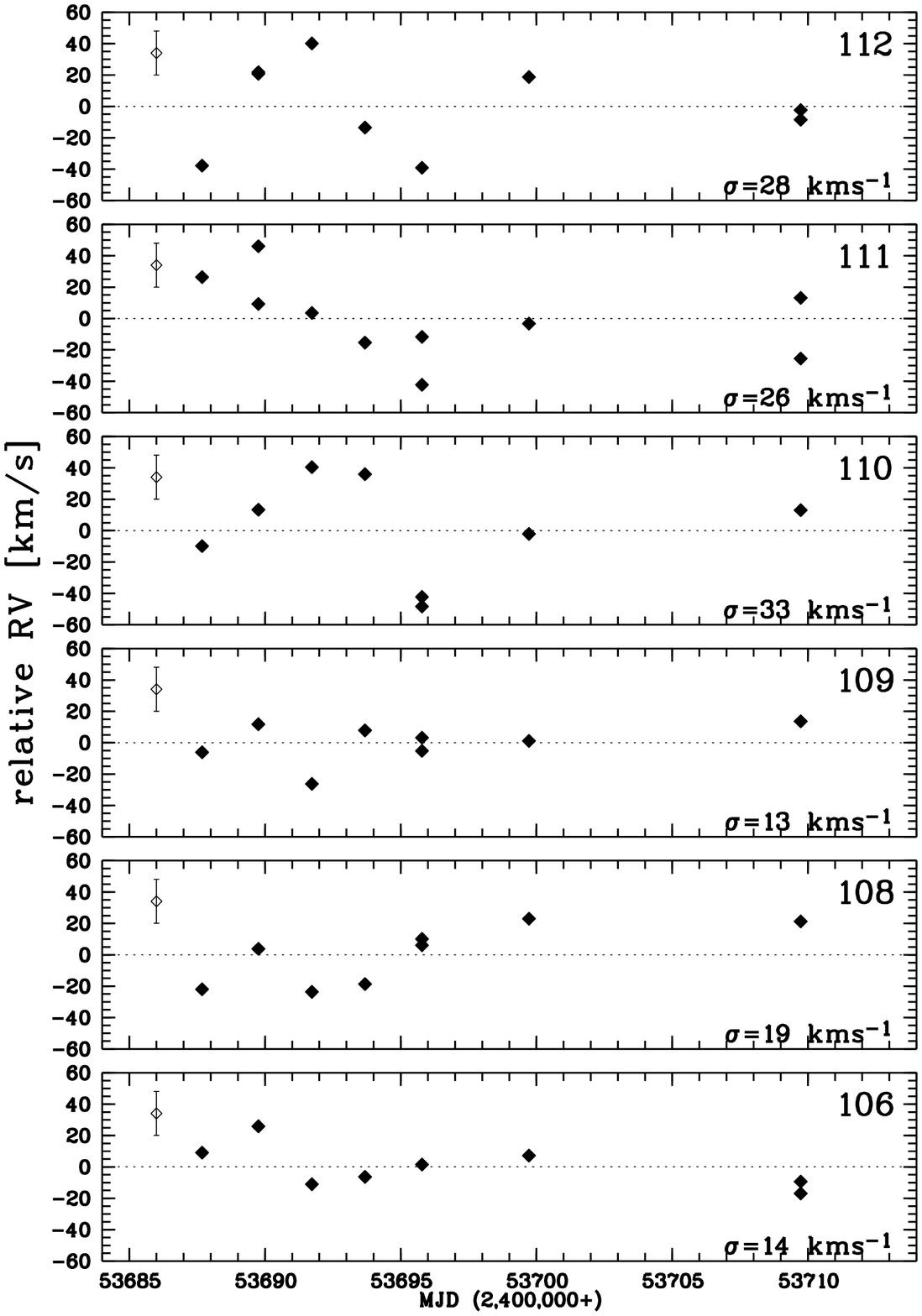}\hfill
\includegraphics[width=85mm,angle=0,trim= 5 0 25
50,clip]{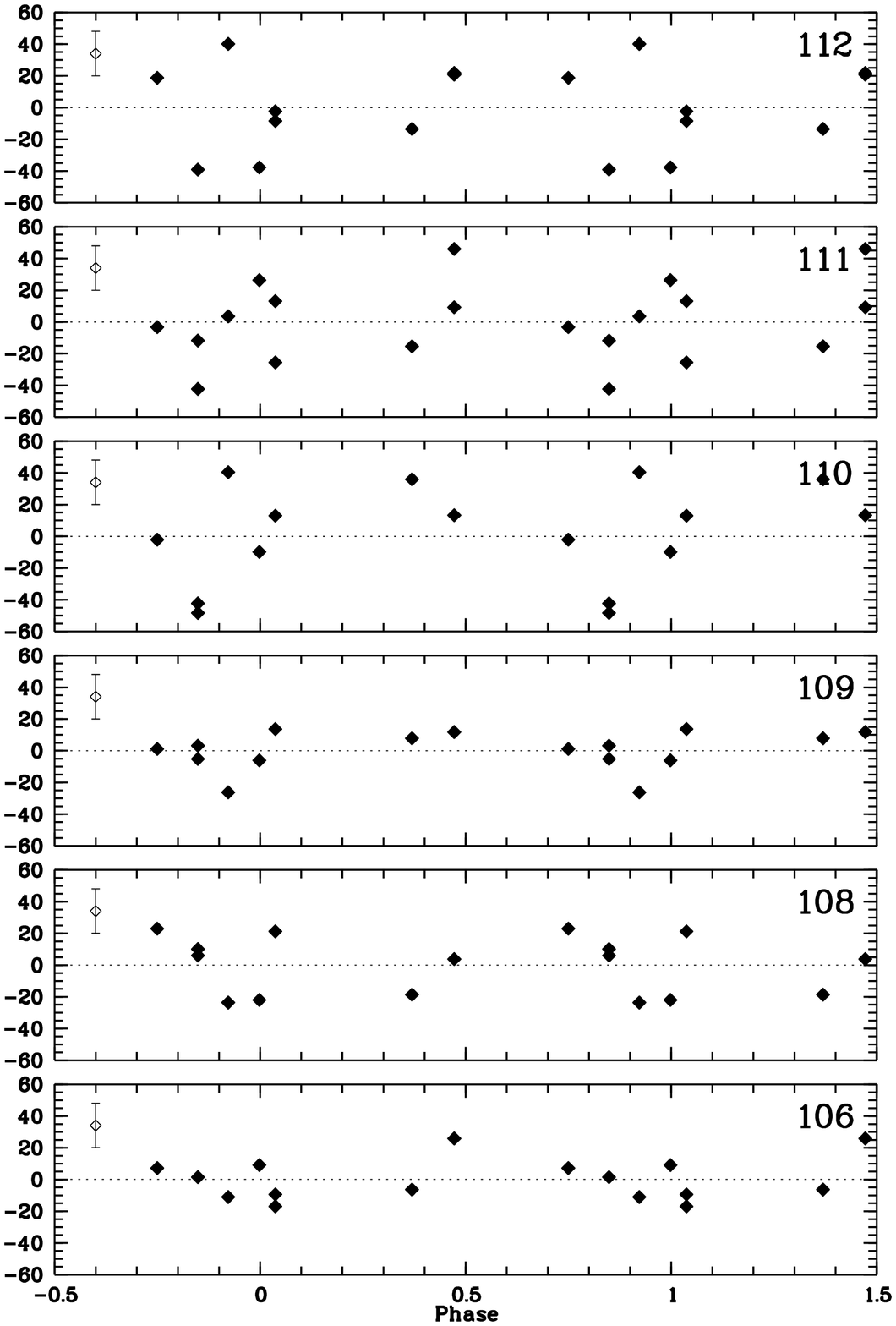}
\end{minipage}
\caption{Relative radial velocities as obtained from
cross-correlation, normalized so that the average $\overline{RV} = 0$
kms$^{-1}$. Error bars ($\pm \sigma$) shown in the upper left of each
panel are adopted from the reference stars, $\sigma_{\rm RV,all} = 14$
kms$^{-1}$. Zero velocity is shown by the dashed line. Left graphs
show RVs plotted against time, right graphs show the data folded into
the phase corresponding to the period of P = 4.377 days of the
putative binary (with arbitrary zero phase). BAT99 numbers are given
in the upper right corner of each panel. No coherent RV variation can
be seen in the data, with the possible exception or BAT99-112 (R136c;
top left row). See text for more details.}
\label{rvtime}
\end{figure*}

To identify the variable stars in our sample, relative radial
velocities (RVs) were measured by cross-correlation, limited to the
region from 2.134 to 2.214 $\mu$m which comprises the two strongest
emission lines, Br$\gamma$/He\textsc{i} $\lambda 2.166 \mu$m and
He\textsc{ii} $\lambda 2.188 \mu$m. The iterative approach of
\citet{S08b} was followed: A high-S/N mean spectrum of the time series
acted as cross-correlation template for each star. Then, all spectra
of the time series were shifted by their respective RVs, to generate a
better mean template spectrum, and the cross-correlation was repeated
with the new template. Figure \ref{rvtime} shows time-plots of these
relative RVs for each star.

Even without this iteration scheme, the resulting RV scatter of each
individual star, $\sigma_{\rm RV,xcorr}$ is surprisingly small; final
values are listed in Table \ref{scatters}. Following the approach of
\citet{S08b} to derive \emph{a posteriori} error bars, we proceeded to
construct a RV reference star by using the least variable stars in our
sample, the three WN5h stars in R136a. For each of these stars,
relative RVs obtained by cross-correlation were normalized to
$\overline{RV} = 0$ to correct for different velocities of the
cross-correlation template, and combined. For the resulting 33 data
points in total, the \emph{overall} standard deviation $\sigma_{\rm
RV,all} = 14$ kms$^{-1}$ was obtained. This value was adopted as
measurement error for each individual data point.

To determine whether or not the RV variability of a given star is
statistically significant, we adopt $\sqrt{2} \times \sigma_{\rm
RV,all} = \sigma_{\rm cut}$ as a rough estimate of the threshold above
which a star can be considered to be variable at the 99.9\% level (see
\citealt{S08b} for a more detailed discussion). We obtain $\sigma_{\rm
cut} = 20$ kms$^{-1}$ from our three reference stars, and find that
R136a5, R136b, and R136c exceed this threshold. However, the large
scatter for the first two stars can be explained by very weak lines
and thus relatively lower S/N, which greatly affects the precision of
the RV measurement. In contrast to this, R136c is a stronger-lined
WN5h star, and well isolated, hence its large RV scatter is likely
intrinsic.

Including R136c in our RV reference would yield a somewhat larger
error, $\sigma_{\rm RV,all} = 19$ kms$^{-1}$, and accordingly
$\sigma_{\rm cut} = 27$ kms$^{-1}$. Hence, R136c would have exceeded
the cut-off value in any case (if just), thus regarded as variable,
and been removed from the reference group.

A period search was carried out on the RV data of R136c, constrained
to the range from 2 to 44 days. The lower limit of this range is set
by the mean sampling frequency, while the upper limit is twice the
total time covered by the observations, 22 days. However, due to the
low level of variability, no significant (at or above $3\sigma$)
period can be found. Furthermore, the scarcity of data points leads to
many empty phase bins if a phase-dispersion method is applied. Thus,
more and better data are required to confirm the variability of R136c,
and eventually its binary status.

\begin{figure}
\includegraphics[width=60mm,angle=-90,trim= 0 15 10 38,clip]{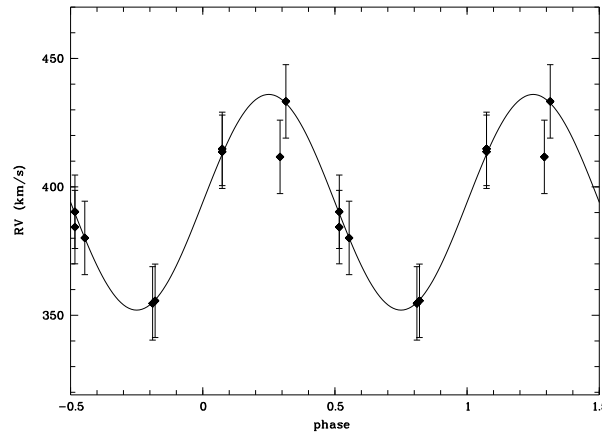}
\caption{Orbital solution for BAT99-112 (R136c) if a sine wave (i.e. a
circular orbit) is forced to the RV data obtained through
cross-correlation.}
\label{sinefit112}
\end{figure}

However, if a sine wave is fitted to the RV data of R136c
(i.e. assuming zero orbital eccentricity), the (marginally) best fit
is obtained for a period $P = 8.2$ days, with an amplitude $K = 42$
kms$^{-1}$ (see Figure \ref{sinefit112}). For this solution,
$\sigma(\rm o-c) = 8$ kms$^{-1}$, which is unrealistically small, and
essentially produced by only one data point (around $\varphi = 0.3$),
which does not lie on the curve. If this point is excluded from the
fit, then $\sigma(\rm o-c) \sim 2$ kms$^{-1}$, which is an indication
for an ill-constrained fit (effectively six independent data points
for four parameters) rather than real.  While it is not impossible
that SINFONI delivers such good data, it sheds some doubt onto the
reality of the orbital solution. We therefore consider R136c as a
marginal case for supporting a binary, subject to future confirmation.

\subsection{Photometry}

Crowded-field, quasi-white light photometry was extracted from each of
the STIS frames using the \textsc{daophot ii} package with the
additional \textsc{allstar} and \textsc{allframe} subroutines
({\citealt{Stetson92}; \citealt{Stetson94}). The point-spread function
(PSF) was iteratively fitted using a Moffat function ($\beta=1.5$)
whose full-width at half-maximum (FWHM) was typically $\sim1.7$
pixels. The fitting radius around stars was set to 4 pixels and the
extraction radius of the aperture was 1.5 FWHM $\sim2.5$ pixels. Sky
was determined from an annulus between 7 and 9 pixels from the center
of the star. In order to separate all the stars in the crowded center
of the cluster, profile-fitting was done iteratively. That way, we
were able to obtain reliable photometry of BAT99-108, 109 and 100, as
well as 3 other fainter stars in the crowded field around BAT99-109.

To carry out differential photometry, a single adjustment to the
photometric zero point was applied as in \citet{Massey02}. For each
star, the dispersion around the time-averaged magnitude is plotted
against its instrumental magnitudes (Figure \ref{sigmamag}). Indicated
are both our program stars and the binaries and variable stars
reported by \citet{Massey02}. Since the exposures were very short
($\sim1.3$\,sec), our target stars, which are the brightest objects in
the field, remain well below the non-linearity limit.

\begin{figure*}
\includegraphics[width=135mm,angle=0,trim= 20 0 0 0,clip]{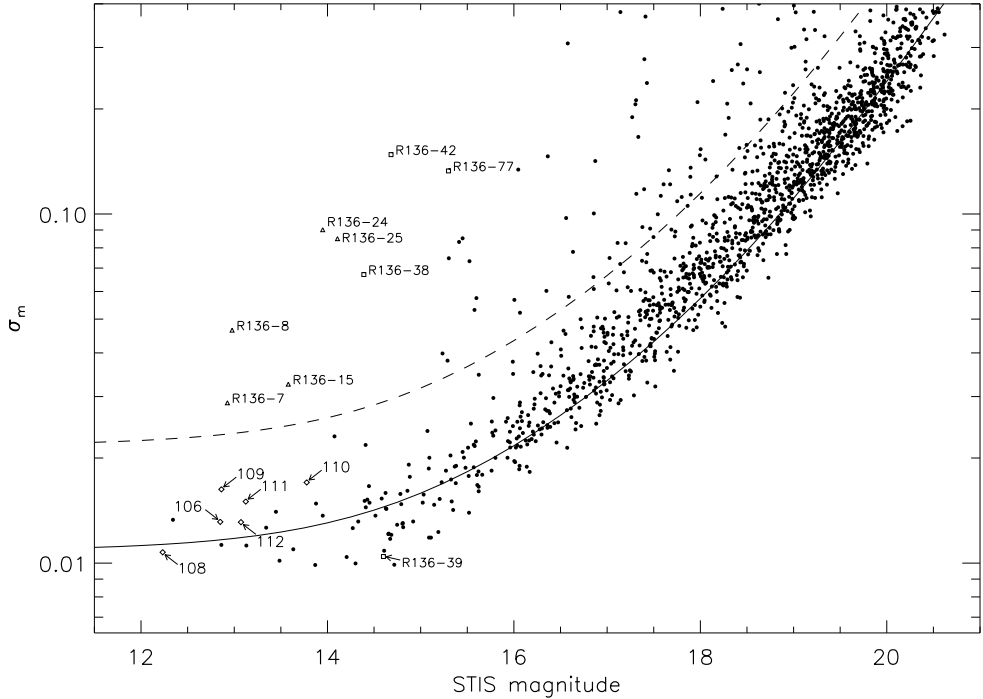}
\caption{Photometric dispersion of the times-series of each source in
the STIS field-of-view, plotted against its instrumental
magnitude. The solid line is the $1\sigma$ curve, the dashed line the
$2\sigma$ curve according to the model described in the text. Our
target stars are identified by their respective BAT99 number. We have
also identified eclipsing binaries (squares) and other variables
(triangles) that were reported by \citet{Massey02}, using their
nomenclature (R136-nnn).}
\label{sigmamag}
\end{figure*}

These very short exposures times have repercussion on the photometric
error statistics. If there were no exterior noise sources, photometric
accuracy would be simply described by Poissonian statistics, and
related to the stellar intensity I such that $\sigma_{\rm I} =
\sqrt{\rm I}$. In reality, however, read-out noise becomes quickly
very important for the photometric accuracy of fainter stars. The
photometric precision of very bright stars, on the other hand, is
limited by the quality of the flat-field frame. To obtain a reliable
error statistics, we therefore have applied a more detailed noise
model, taking into account the following noise sources: $i$) The
Poisson noise of the stellar intensity I, which is simply $\sigma_{\rm
I}$. $ii$) The read-out noise $\sigma_{\rm R}$, which for STIS is 4
e$^{-}$ per pixel (gain = 1). Note that the mean read-out level
$\langle \rm R \rangle = 0$, since there is no flux associated with a
read-out operation, only an error on the count rate. The extraction
radius of the PSFs are $\sim 2.5$ pixels (see above), hence PSFs cover
a total area of $(2.5)^{2} \pi \sim 20$ pixels; hence the total
read-out noise sums to $4 \times \sqrt{20} = 18$ e$^{-}$. $iii$) The
noise $\sigma_{\rm F}$ of the flat-field frame, which we assume to be
normalized to unity prior to dividing the science exposures by it, so
that the mean flat-field factor $\langle \rm F \rangle = 1$.

Since the \emph{HST} is in space and exposures are too short for a
significant stray-light contribution, there is no sky background and
its associated noise that need to be considered. We further note that
the flat-field noise is multiplicative (it is proportional to the
intensity recorded by a pixel), whereas the read-out noise is purely
additive. Hence, the photometric precision for bright stars is limited
by the quality of the flat-field frame, whereas faint stars are more
affected by read-out noise.

Thus, from the usual relation

\begin{eqnarray*}
m = m_{0} -2.5 \log(\rm F \times \rm I + \rm R),
\end{eqnarray*}

\noindent
where $m_{0}$ is a magnitude zero point, error propagation for
independent values of F, I, and R then yields the total error on the
measured magnitude,

\begin{eqnarray*}
\sigma^{2}_{\rm m} = \left( \frac{2.5 \log e}{\rm F \times \rm I + \rm R} \right)^{2} + \left[ (\rm I \times \sigma_{\rm F})^{2} \times (\rm F \times \sigma_{\rm I})^{2} + \sigma^{2}_{\ R} \right]\end{eqnarray*}

\vspace{0.5cm}
\noindent
(with $e$ the Euler number), which reduces to

\begin{eqnarray*}
\sigma^{2}_{\rm m} = \left( \frac{2.5 \log e}{\rm I} \right)^{2} \times \left[ (\rm I \times \sigma_{\rm F})^{2} + \rm I + \sigma^{2}_{\ R} \right].
\end{eqnarray*}

\vspace{0.5cm}
\noindent
Using $\sigma_{\rm F} = 0.01$ (which is consistent with the typical
S/N of a flat-field frame), and $m_{0} = 24.9$, this noise model fits
the data very well. Both the $1\sigma$ and $2\sigma$ curves, based on
this model, are shown in Figure \ref{sigmamag}. As can be seen, even
the brightest stars in our sample do not yet reach the noise floor
that is determined by the flat-field noise.

As can clearly be seen, our target stars do not display any
significant photometric variability; in fact the dispersion around
their respective mean magnitudes is very small (see Table
\ref{allphot} for individual values). Since we were perfectly able to
identify the eclipsing binaries and other variable stars reported by
\citet{Massey02}, we have to conclude that our target stars appear to
be photometrically constant. This is especially relevant for BAT99-112
(R136c), which despite its short spectroscopic period does not show
any indications of phase-dependent variability.

\begin{table}
\begin{center}
\caption{Mean instrumental magnitudes (through the long-pass filter)
and standard deviation around the mean for our target stars as
obtained from \emph{HST}-STIS images using \textsc{daophot ii}.}
\label{allphot}
\begin{tabular}{clcc}
\hline
BAT99   & other   & STISmag &$\sigma_{\rm phot}$ \\
        & name    & \multicolumn{2}{c}{(mag)}       \\
\hline
106     & R136a3  &  12.850 & 0.013    \\
108     & R136a1  &  12.233 & 0.011    \\
109     & R136a2  &  12.863 & 0.016    \\
110     & R136a5  &  13.778 & 0.017    \\
111     & R136b   &  13.124 & 0.015    \\
112     & R136c   &  13.072 & 0.013    \\
\hline
\end{tabular}
\end{center}
\end{table}

\section{Discussion}

With our data, we cannot confirm the existence of the 4.377-day binary
that was reported in optical emission lines by \citet{MoffSegg83} and
\citet{Moff85}. Moreover, we cannot clearly identify any short-period
binary among our six target stars; there only is one binary candidate,
BAT99-112 (R136c). This star was already known to be a very bright
X-ray source. Several studies have reported a very high X-ray
luminosity of $L_{\rm X} \ga \rm 8.5 \times 10^{34}$ ergs s$^{-1}$ for
BAT99-112 (\citealt{Portegies02}; \citealt{Townsley06};
\citealt{Guerrero08}), in fact the second-brightest object in the
greater R136 area after BAT99-116 (Mk 42). It is very likely that both
stars are colliding-wind binaries
(cf. \citealt{Usov92}). Interestingly, \citet{S08b} reported
significant RV variability for BAT99-116 as well, but were unable to
establish a periodicity. While it is thus very likely that both are
long-period systems, we note that short-period systems, i.e. with a
small orbital separation, are not ruled out by the high X-ray
luminosities. NGC3603-C is an 8.9-day binary with one of the highest
X-ray fluxes known among all WR stars (\citealt{Moff02};
\citealt{S08a}), so it seems that self-absorption of X-ray photons in
the dense WN wind is not too much of a problem even in relatively
close systems.

Thus, the following question arises: If one or more of our target
stars in R136 are indeed short-period binaries, could their
non-detection be the result of low orbital inclination angles? After
all, very low inclination angles could comfortably explain why neither
clear RV nor photometric variability (i.e., eclipses or ellipsoidal
variations) are found, despite any possible enhancement of
short-period systems (see above). To investigate this in more detail,
we have calculated the RV scatter $\sigma_{\rm RV,WN} = K / \sqrt{2}$
of the primary (WN-type) component that would be expected from the
continuous sampling of a circular binary with orbital periods ranging
from 1 to 100 days. In this case study, the masses of the two
components have been fixed to 90 $M_{\odot}$ for the WN5h primary and
30 M$_{\odot}$ for the presumed O-type secondary, i.e. the system has
a total mass of 120 M$_{\odot}$ (the exact value is not critical), and
a mass ratio $q = M_{1}/M_{2} = 3$. The large $q$ value is reasonably
pessimistic, given that such large ratios are observed even in very
massive systems: In many reported cases, the companion is too faint to
be (easily) detected (e.g. \citealt{Schweick99}; Schnurr et
al. 2008a,b; \citealt{S09a}).

The expected $\sigma_{\rm RV,WN}$ value for different values for the
inclination angle, $i = 15^{\circ}$ to $90^{\circ}$ in steps of
$15^{\circ}$, are shown in Figure \ref{detectionlimits}. Also
indicated are, with two vertical, dotted lines, the position of the
expected 4.4-day binary, and the longest period we can reasonably hope
to find in our data set, twice the time coverage of our observations,
i.e. $\sim$44 days. The dash-dotted, horizontal line marks the
observed RV scatter for BAT99-112, $\sigma_{\rm RV,112} = 28$
kms$^{-1}$.

From Figure \ref{detectionlimits} it becomes clear that if BAT99-112
were indeed a short-period binary, it would have to be seen under a
very low inclination angle, $i \la 15^{\circ}$. However, in a sample
with randomly distributed inclination angles, there would be

\noindent
\begin{eqnarray*}
\frac{\int_{15^{\circ}}^{90^{\circ}} \sin{i} \, di}{\int_{0^{\circ}}^{15^{\circ}} \sin{i} \, di} = \frac{\cos{15^{\circ}}}{1- \cos{15^{\circ}}} \sim 28
\end{eqnarray*}

\noindent
times more binaries between $15^{\circ}$ and $90^{\circ}$ than 
between $0^{\circ}$ and $15^{\circ}$; thus, it is very unlikely
that there is a significant number of such low-inclination systems in
our sample.

\begin{figure}
\includegraphics[width=60mm,angle=-90,trim= 0 0 0 0,clip]{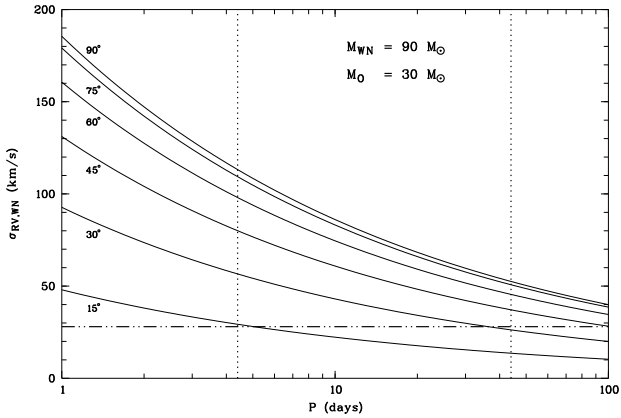}
\caption{RV scatter $\sigma_{\rm RV} = K / \sqrt{2}$ of the WN
component in a fictuous, continuously sampled, circular binary, with
masses fixed at 90 M$_{\odot}$ for the WN5h primary and 30 $M_{\odot}$
for the O-type secondary, as a function of the orbital period of the
system. Curves show different values of the orbital inclination angle,
$i = 90, 75, 60, 45, 30, 15^{\circ}$, from top. The dot-dashed,
horizontal line indicates the measured scatter for R136c (BAT99-112),
$\sigma_{\rm RV} = 28$ kms$^{-1}$. The two dotted, vertical lines
indicate the period of the putative binary, 4.4 days, and the upper
limit of periods we can reasonably hope to detect in our 22-day long
campaign, 44 days.}
\label{detectionlimits}
\end{figure}

We note here that \citet{MoffSegg83} and \citet{Moff85} had reported a
RV amplitude of $K \sim 37$ kms$^{-1}$ for R136a, which seems
surprisingly consistent with $\sigma_{\rm RV,112} \times \sqrt{2} =
40$ kms$^{-1}$. However, \citet{MoffSegg83} and \citet{Moff85} only
had integrated, i.e. spatially unresolved spectroscopy, and they
reported the binary to be member of R136a, i.e. one of the three WN5h
stars. (The fourth star, BAT99-110, is O3f/WN6 and hence too faint
both in magnitude and emission-line strength to significantly have
contributed to the integrated signal.)  Our SINFONI observations,
however, have individually resolved the stars; hence one would expect
an RV amplitude at least three times higher, i.e. $K \sim 120$
kms$^{-1}$. Even if one allows for the fact that due to non-continuous
sampling, the measured $\sigma_{\rm RV}$ could be somewhat smaller,
such large RV scatter is simply not seen among the WN5h stars in
R136a.

Hence, we have to conclude that the large scatter in Moffat \&
Seggewiss's (1983) and Moffat et al.'s (1985) data was not due to the
binary motion of one of the target stars, but of different origin. A
possible explanation could be that during the observations, the slit
was not located at exactly the same position each time, e.g. due to
difficulties introduced by variable seeing conditions, and that thus
the light took different optical paths through the instrument. This
could lead to an increased RV scatter, while well-isolated single
stars are much less affected by this problem (see
e.g. \citealt{S08b}). It thus seems that with the exception of the
8.2-day (but possibly longer-period) binary candidate BAT99-112, there
is no binary among our six program stars.

For BAT99-111 (R136b), this result does not come unexpectedly:
\citet{CroDess98} re-classified this star as WN9ha, and there seems to
be a dichotomy between WN6,7 and WN8,9 stars when it comes to their
binary status: No WN8,9 star is known to reside in a short ($P \la
200$ days) WR+O binary system (Milky Way: \citealt{M89}; LMC:
\citealt{S08b}). Also, if BAT99-111 is indeed a WN9ha star, then it is
a descendent of an O8Iaf star (\citealt{CroBo97}), i.e. less massive
but slightly more evolved than the O3f/WN6 and WN5h stars in
R136. This could be an indication that BAT99-111 is not a member of
R136a's relatively unevolved stellar generation, but rather from an
older population of massive stars outside R136, and simply located in
the line of sight.

If, on the other hand, BAT99-111 is indeed an O4Iaf$^{+}$ star, as
classified by \citet{MassHunt98}, then both the spectral type and the
fainter visual magnitude are consistent with BAT99-111 being a less
extreme version of the brighter O3f/WN6 stars in and around R136, and
a member of R136a after all. Unfortunately, our data do not allow us
to settle this classification issue.

For the other five, less evolved objects in our sample, in particular
for the three very crowded WN5h stars in the core of R136a, the
null-result is somewhat surprising. We had expected to find at least
one short-period binary in R136a, similar to the situation in its
almost perfect, Galactic twin NGC3603, where \citet{S08a} reported
that of the three central WNh stars, two are binaries with $P < 10$
days. Also, \citet{S08b} report that among the nine O3f/WN6 and WN5-7h
stars in 30 Dor, but outside R136, only two ($\sim$20\%) are confirmed
spectroscopic binaries. This value (at least) should have been
recovered in the core of R136 as well, since in the present study, we
are using data of almost identical quality.

Considering the small number of studied stars, however, our
non-detection of systems is fully consistent with a frequency of
$\sim$20\% of binaries with $P < 10$ days. In fact, combining our
results with that of \citet{S08b}, we can confirm that the
short-period binary frequency among very massive stars around R136
within 30 Dor is not very high: Out of the total of fifteen O3f/WN6
and WN5-7h stars now studied, only two are confirmed binaries with $P
< 10$ days, i.e. 2/15 = 13\%. If we include the candidate BAT99-112
(R136c) as a positive detection, we obtain 3/15 = 20\% binary
frequency. If the distribution of orbital periods were flat in log\,P
(or proportional to 1/P), something which is usually assumed, we would
find 13\% (20\%) binaries in each period bin [1;10], [10;1000], and
[100;1000], i.e. the total frequency of binaries with orbital periods
shorter than 1000 days thus would be 40\% (60\%).

Very recently, \citet{Bosch09} have reported on the results of a
spectroscopic monitoring campaign of absorption-line OB stars in the
30 Dor region; for 52 stars, six to seven spectra per star were
obtained over a time span of $\sim$500 days. While this small number
of spectra does not allow one to establish an orbital solution,
\citet{Bosch09} were able to identify binaries from their
significantly large RV scatter compared to single stars and binaries
with much longer periods; among the 52 stars monitored, 25 stars
showed RV variability. If we suppose that with their method,
\citet{Bosch09} were able to detect binaries with periods up to
$\sim$1000 days (i.e. twice the covered time span), the binary
frequency is thus $\sim48$\%. Taking into account all statistical
errors, our null-result for $P < 10$ days reported here is perfectly
consistent with this value. Remarkably, Monte Carlo simulations by
\citet{Bosch09} report that their observational results are consistent
with a binary frequency of 100\%. Although our observations were only
designed to identify the reported 4.4-day culprit, and we have not yet
enough data points to identify systems much beyond $P \sim 40$ days,
we can rule out such a high binary frequency for our stars with strong
winds, because only BAT99-112 shows an X-ray excess.

\section{Summary and Conclusion}

We have obtained for the first time, spatially-resolved, repeated,
low-spectral resolution, near-infrared spectroscopy of the most
luminous stars in the core of R136 in an attempt to identify the
4.377-day binary that was reported by \citet{MoffSegg83} and
\citet{Moff85}. Additional archival \emph{HST}-STIS imaging was used
to extract photometry of our target stars.

For none of the studied stars could significant photometric
variability be found. Furthermore, we cannot confirm the presence of a
binary system with the reported 4.4-day period, nor do we identify any
other short-period ($P \la 44$ days) system among the most luminous
stars in R136. One star, however, BAT99-112 (R136c), shows small,
marginally significant RV scatter. A forced sine-fit to the data
yields a best period $P = 8.2$ days. While it is not entirely
impossible, it is not very likely to see a binary under very low
inclination angle ($i \la 15^{\circ}$), which would be required to
reconcile BAT99-112's 8.2d binary nature with the observed, low RV
scatter. Thus, it is more probable that BAT99-112 is a longer-period
system that the limited time coverage of our observations could not
detect, as is suggested by the fact that the star is one of the
brightest X-ray sources among all known WR stars in the LMC.

To settle this issue, long-term monitoring of BAT99-112 will be
required, but the possibility that it is indeed binary clearly
harbors the potential to weigh one of the most massive stars known,
and to obtain a clearer picture of the very upper main sequence in
high-density, high-mass environments like R136.

The question as to why R136 appears to have a dearth in its
short-period binary content, compared to its Galactic counterpart NGC
3603, remains to be explained, if not simply due to small numbers.  A
key difference, though, is the fact that NGC 3603 is not surrounded by
a massive stellar halo as is R136 (i.e. 30 Dor).  Another difference
is the factor-of-two lower metallicity in R136.  How these or other
as-yet unknown factors play out remains a mystery. In fact,
ultimately, long-term monitoring of all the luminous stars in R136 is
necessary to check for long-period systems.

\section*{Acknowledgements}

OS is grateful for financial support by PPARC/STFC. ANC is grateful to
Peter Stetson for invaluable help with the \textsc{daophot} software
package. AFJM and NSL are grateful for financial aid to NSERC (Canada)
and FQRNT (Qu\'ebec).




\label{lastpage}

\end{document}